\documentclass[%
 amsmath,amssymb,
 aps,
]{revtex4-1}

\usepackage{graphicx}
\usepackage{dcolumn}
\usepackage{bm}
\usepackage{hyperref}

\usepackage{float}

\usepackage{enumerate}

\begin{document}

\preprint{APS/123-QED}

\title{Non-standard scaling law of fluctuations in finite-size systems of globally coupled oscillators}

\author{Isao Nishikawa$^{1}$, Gouhei Tanaka$^{1,2}$, and Kazuyuki Aihara$^{1,3}$}
\affiliation{
$^1$Institute of Industrial Science, The University of Tokyo, Tokyo 153-8505, Japan
}%
\affiliation{
$^2$School of Engineering, The University of Tokyo, Tokyo 113-8656, Japan
}%
\affiliation{
$^3$Department of Mathematical Informatics, Graduate School of Information Science and Technology, The University of Tokyo, 7-3-1 Hongo, Bunkyo-ku, Tokyo 113-8656, Japan
}%




\date{\today}

\begin{abstract}

Universal scaling laws form one of the central issues in physics.
A non-standard scaling law or a breakdown of a standard scaling law, on the other hand, can often
lead to the finding of a new universality class in physical systems.
Recently, we found that a statistical quantity related to fluctuations
follows a non-standard scaling law with respect to system size
in a synchronized state of globally coupled non-identical phase oscillators
[Nishikawa et al., Chaos $\boldsymbol{22}$, 013133 (2012)].
However, it is still unclear how widely this non-standard scaling law is observed.
In the present paper,
we discuss the conditions required for the unusual scaling law in globally coupled oscillator systems,
and we validate the conditions by numerical simulations of several different models.

\begin{description}
\item[PACS numbers]
05.45.Xt, 05.50.+q, 05.40.-a, 05.60.-k

\end{description}
\end{abstract}

\pacs{Valid PACS appear here}
\maketitle



\section{Introduction}

One of the essential studies of physics is to reveal universal phenomena in nature \cite{Nishimori}.
Synchronization is a universal concept in nonlinear science that is widely used
to explain coherent behavior in a system of interacting elements \cite{PikovskyBook}
such as a reaction--diffusion system of chemical oscillators \cite{Oertzen},
an array of electrical units \cite{Barone},
spontaneous luminescence of fireflies \cite{Buck},
and a network of rhythmic biological components \cite{Glass}.
A synchronization transition in a system of $N$ coupled oscillators takes place when the coupling strength surpasses a critical value at which an order parameter changes from zero to a non-zero value.
Universality of synchronization phenomena can be partially confirmed by the fact
that the scaling property of the order parameter in the vicinity of the critical synchronization transition point does not depend on the details of the system elements \cite{PikovskyBook,KuramotoBook,Kuramotomodel,Sakaguchi2,Daido2,Crawford2,Chiba2,Daido,Pikovsky4,Hong,Hong3,Hildebrand,Buice,Nishikawa,Nishikawa2}. 
In the thermodynamic limit $N \rightarrow  \infty $,
the study of the universal scaling law with respect to the coupling strength
has been a topic of primary interest \cite{Daido2,Crawford2}.
On the other hand, many studies have also focused on
the universal scaling law with respect to finite system size $N$.
In fact, it is known that the variance of the order parameter
generally decays with $N$ in the order of $1/N$ independently of the details of the system elements \cite{Daido,Hildebrand,Buice,Nishikawa}.
Such a scaling law is regarded as standard
because it applies to a wide range of models.
In the present paper,
we focus on a non-standard scaling law that may provide an opportunity to determine a new universality class.

To illustrate a difference between standard and non-standard scaling laws,
we consider the globally coupled phase oscillator model \cite{KuramotoBook}, described as follows:
\begin{align}
\label{phasemodel}
\dot \theta_j = \omega_j + \frac{K}{N} \sum _{k=1} ^N h(\theta_k - \theta _j)  \ \mathrm{for} \ j=1,...,N,
\end{align}
where $\theta _j$ represents the phase of the $j$th oscillator,
$\omega_j$ represents the natural frequency of the $j$th oscillator, $K>0$ represents the coupling strength,
and $h$ denotes the coupling function.
In the Kuramoto model where $h(x) = \sin(x)$ \cite{KuramotoBook,Kuramotomodel},
the oscillators are synchronized when the coupling strength $K$ is sufficiently large.
Synchronization transitions
have been extensively studied with regard to their analogy to the second-order phase transition \cite{KuramotoBook,Kuramotomodel}.
A measure of synchrony is given by the order parameter $R(t)$ as follows \cite{KuramotoBook}:
\begin{align}
\label{R}
R(t) \equiv \frac {1}{N} \left|\sum _{j=1} ^N \exp(2\pi i \theta _j)\right|.
\end{align}
A synchronization transition is characterized by a change in the order parameter $R(t)$ from almost zero to a finite value with an increase in the coupling strength $K$.
In the thermodynamic limit $N\rightarrow \infty $,
the stationary state of $R(t)$ in the incoherent (desynchronized) regime bifurcates at the critical coupling strength $K=K_c$,
above which the oscillators are coherent (synchronized).
When $N$ is finite, the effect of an increase in $N$ on fluctuations can be different between the incoherent and coherent regimes \cite{Nishimori,Nishikawa}.
To illustrate this, we introduce the diffusion coefficient of the time integral of $R(t)$, characterizing long-term fluctuations of $R(t)$ in systems of finite size \cite{Nishikawa}.
We consider the variance of the time integral $\int _0 ^t R(s) ds$, defined as follows:
\begin{align}
\label{sigma}
\sigma^2(t) \equiv N \left[  \lim _{T \rightarrow  \infty }  \frac{1}{T}  \int _{0} ^{T}   \left( \int _{t_0} ^{t+t_0} R(s) {\rm d}s - \langle R \rangle _t  t \right) ^2  {\rm d}t_0    \right] _s,
\end{align}
where $\langle \cdot \rangle _t$ represents the long-time average,
and $[ \cdot ] _s$ denotes the sample average over different realizations of $\omega _j $ that are chosen from a certain distribution. 
For the variance in Eq.~$(\ref{sigma})$, the following diffusion law holds \cite{Nishikawa}:
\begin{align}
\label{D}
D \equiv \lim _{t\rightarrow \infty } \sigma^2(t) /  2t,
\end{align}
which represents the deviation of $\int _0 ^t R(s)ds$ from its mean value $\langle R \rangle _t  t$ on the sample average.
The standard scaling law of $D$ with respect to system size $N$ is described as follows \cite{Nishikawa}:
\begin{align}
\label{normalDN} D \sim O(1).
\end{align}
Namely, the value of $D$ is almost constant independently of the system size $N$.
We numerically confirmed that
this scaling law is observed in the Ising model \cite{Nishimori},
the probabilistic cellular automaton model \cite{Nishimori,Bagnoli,Takeuchi},
and the random neural network model \cite{Toyoizumi}.
In the previous study \cite{Nishikawa},
we also confirmed that
the standard scaling law holds in the incoherent regime of the globally coupled phase oscillator model (\ref{phasemodel}).
In our discussion in Ref. \cite{Nishikawa},
we suggested that the standard scaling law can be observed in a wide range of systems.
We also found that, in the coherent regime of system (\ref{phasemodel}),
the asymptotic form of $D$ for large $N$ is represented as
\begin{align}
\label{DN} D \sim O(1/N^a),
\end{align}
where $a > 0$ is a certain constant.
This non-standard scaling law suggests the existence of a new universality class.
However, it is still obscure how broadly this scaling law is observed in coupled oscillator systems.
The aim of this study is to elucidate the validity range of the above non-standard scaling law of $D$.

\section{Necessary condition for non-standard scaling law}

\begin{figure}
\begin{center}
\centering
\begin{tabular}{ll}
\includegraphics[width=5cm,clip]{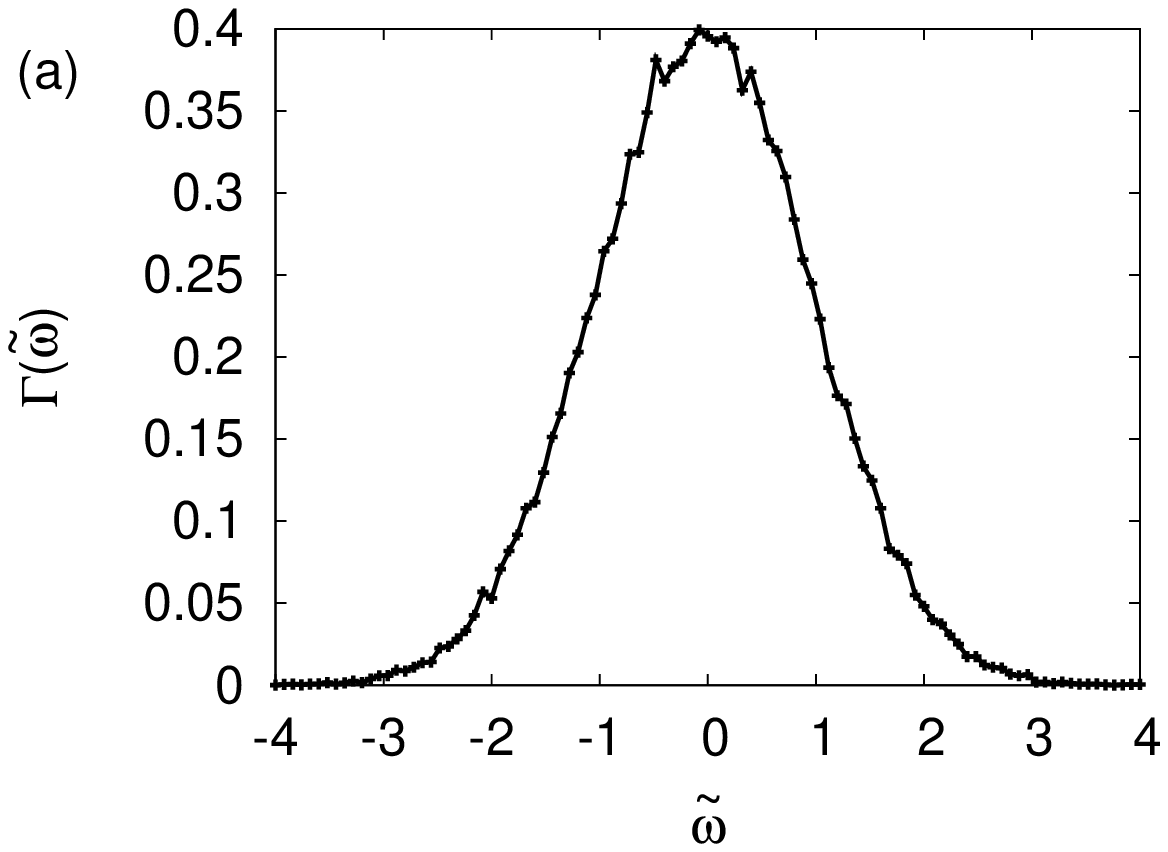}
\includegraphics[width=5cm,clip]{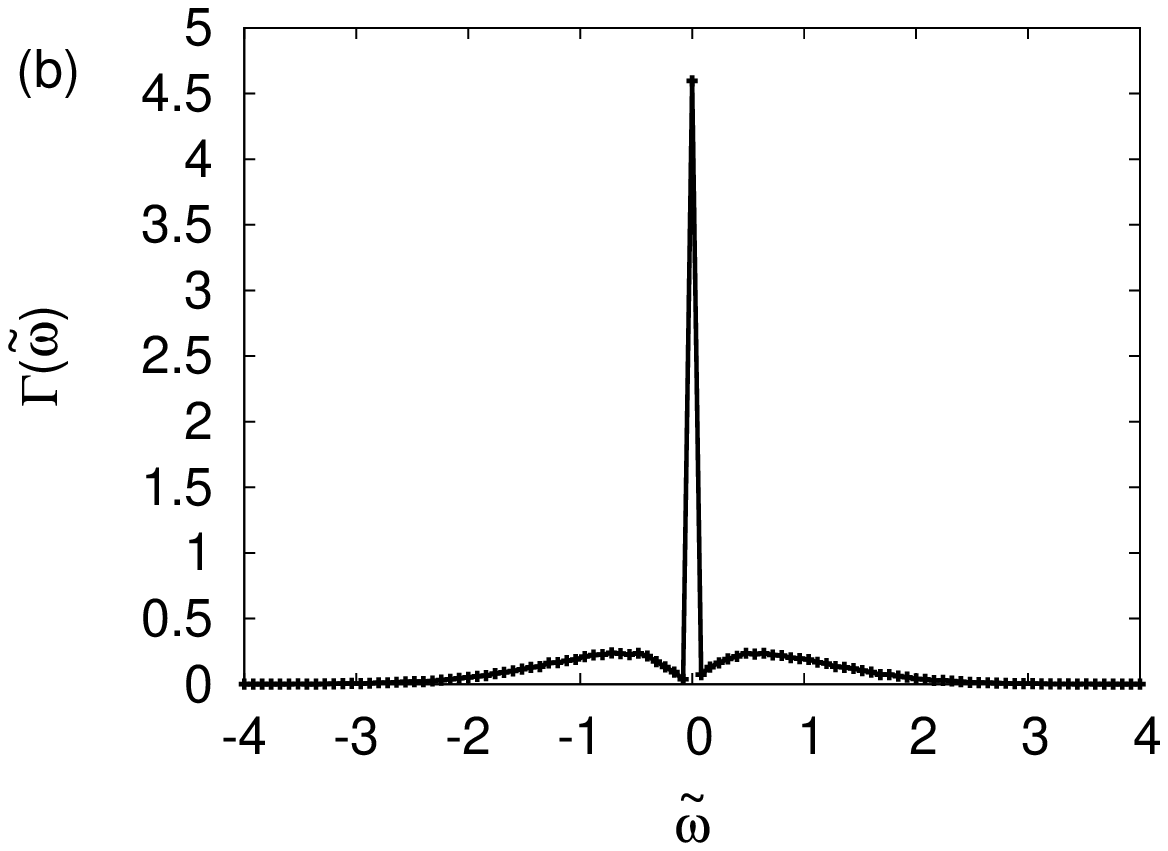}
\end{tabular}
\begin{tabular}{ll}
\includegraphics[width=5cm,clip]{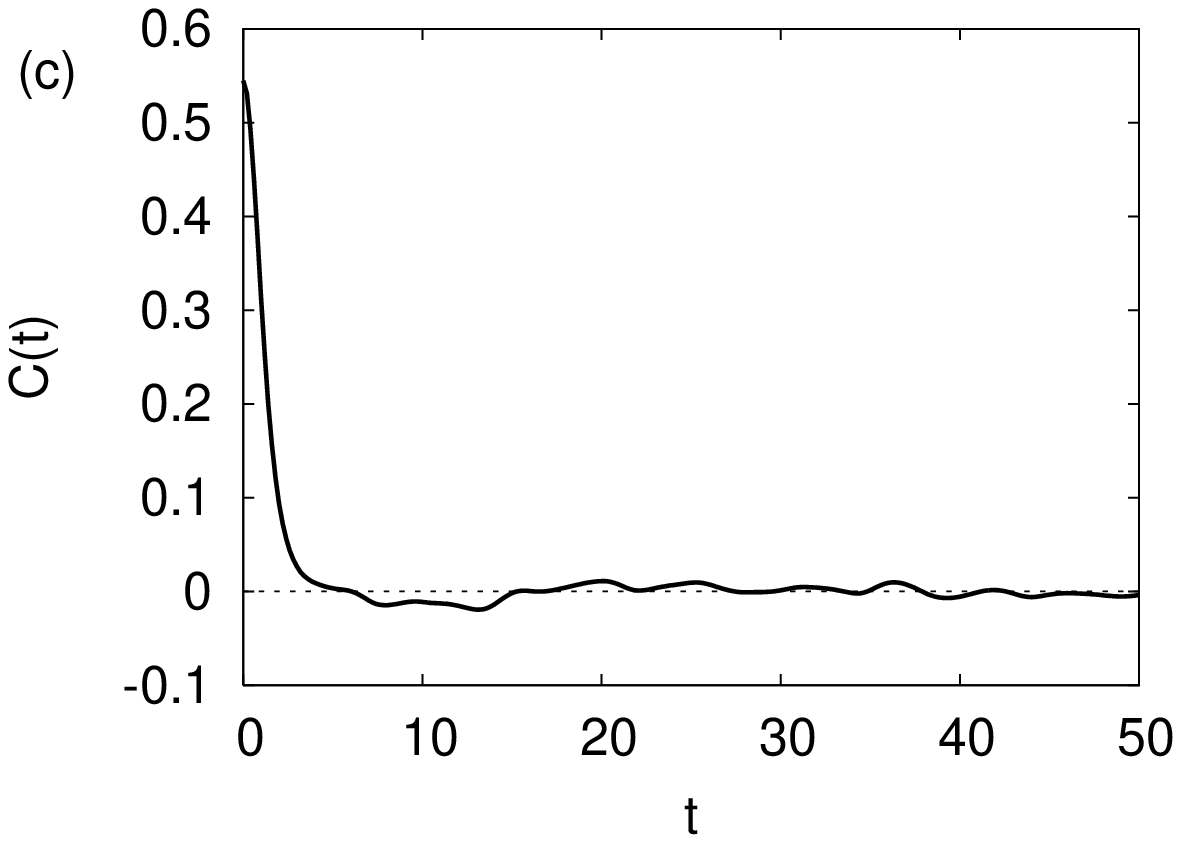}
\includegraphics[width=5cm,clip]{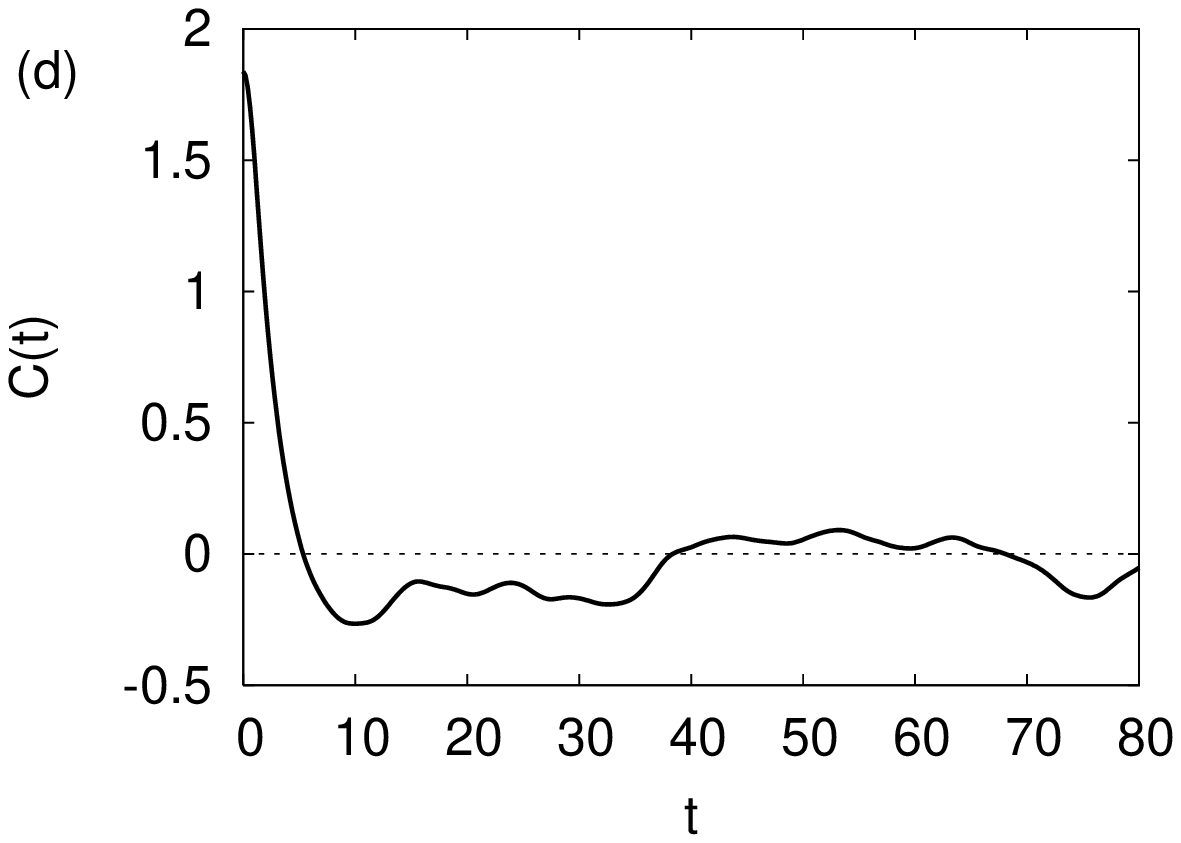}
\end{tabular}
\end{center}
\caption{\label{fig:schematic2}
Numerical results of $\Gamma (\tilde \omega )$ (upper panels) and the time evolution of the correlation function $C(t)$
(lower panels) for the Kuramoto model (Eq.~(\ref{phasemodel}) with $h(x) = \sin (x)$) with $N=64000$.
The natural frequencies $\omega_j$ are randomly drawn from a Gaussian distribution with mean zero and standard deviation one.
(a) and (c) An incoherent state with $K=0.8$. (b) and (d) A coherent state with $K=1.63$}
\end{figure}

Let us consider the mean frequencies $\tilde \omega _j = \lim _{t \rightarrow  \infty } (\theta _j (t) - \theta _j (0))/t$ of the phase variables in the stationary state.
We define by $\Gamma (\tilde \omega )$ the distribution function of the mean frequencies of all the oscillators.
A necessary condition for the scaling law $(\ref{DN})$ in system (\ref{phasemodel}) is given as follows \cite{Nishikawa}:
\begin{align}
\label{criticaleq}
\lim _{\tilde \omega \rightarrow \Omega } \Gamma (\tilde \omega ) = 0  \ \mathrm{with} \ N \rightarrow \infty,
\end{align}
where $\Omega $ denotes the common frequency of the entrained oscillators.
A typical example is illustrated in Figs.~$\ref{fig:schematic2}$(a) and $\ref{fig:schematic2}$(b),
where we observe that Eq.~$(\ref{criticaleq})$ holds in a coherent state whereas it does not hold in an incoherent state \cite{PikovskyBook}.
In Fig.~$\ref{fig:schematic2}$(a),
oscillators with $\tilde \omega \sim 0$ are dominant and oscillators with fast oscillations very few.
In Fig.~$\ref{fig:schematic2}$(b),
we can observe the largest peak corresponding to entrained oscillators
and two other peaks corresponding to non-entrained oscillators
with fast oscillations.
It has been previously shown that
the value of $D$ is mostly determined by the value of $\Gamma (\tilde \omega )$ around $\tilde \omega = \Omega $ \cite{Nishikawa}.
The value of $D$ usually decreases with system size $N$, if Eq.~$(\ref{criticaleq})$ holds.

To understand the difference between the standard and non-standard scaling properties,
we consider the following correlation function of the order parameter:
\begin{align}
\label{C}
C(t) \equiv N \left[  \lim _{T \rightarrow  \infty }  \frac{1}{T}  \int _0 ^{T}     ( R(t+t_0) - \langle R \rangle _t)( R(t_0) - \langle R \rangle _t)   {\rm d}t_0   \right]_s .
\end{align}
Consequently, we can relate $C(t)$ to $D$ as follows \cite{Correlationfunction,Nishikawa}:
\begin{align}
\label{CD}
D = \frac{1}{2} \int _ {-\infty} ^{\infty} C(s)  {\rm d}s.
\end{align}
Therefore, the form of the correlation function $C(t)$ determines the value of $D$.
In the incoherent regime of system (\ref{phasemodel}),
the form of $C(t)$ is almost exponential as shown in Fig.~\ref{fig:schematic2}(c) \cite{Nishikawa,Daido}.
In contrast, in the coherent regime,
the non-entrained oscillators generate fluctuations in the correlation function $C(t)$ around $C(t)=0$
as shown in Fig.~\ref{fig:schematic2}(d) \cite{Nishikawa,Daido}.
Therefore, a characteristic time period with $C(t)<0$ is observed.

\section{Generality of the non-standard scaling law}

Let us discuss what type of coupled oscillator models satisfy the condition $(\ref{criticaleq})$.
It has been reported that this condition is not satisfied in a coupled oscillator model with local coupling \cite{Hong},
one with very non-uniform coupling strength \cite{Daido5},
and one with a large variance in the natural frequencies \cite{Matthews}.
To avoid these cases,
it appears to be necessary that
a model involves global coupling \cite{PikovskyBook,KuramotoBook,Sakaguchi2,Daido2,Daido3},
nearly uniform coupling strength \cite{Aonishi},
and a sufficiently small variance in the natural frequencies \cite{Nagai}.

We examine the existing models that are known to satisfy the condition $(\ref{criticaleq})$
in order to confirm whether the scaling law $(\ref{DN})$ holds or not.
The results of our numerical simulations, presented in the following part, suggest that
the condition $(\ref{criticaleq})$ is not only a necessary but also a sufficient condition for the non-standard scaling law.

First, we investigate the phase oscillator model with time-delayed coupling \cite{Choi}, described as follows:
\begin{align}
\label{phasemodel-timedelay}
\dot \theta_j = \omega_j + \frac{K}{N} \sum _{k=1} ^N \sin (\theta_k (t - \tau ) - \theta _j (t) ),
\end{align}
where $\tau $ represents the time delay.
Equation $(\ref{criticaleq})$ holds for any time delay $\tau $ \cite{Choi}.
Second, we consider the phase oscillator model with inertia \cite{Tanaka} given by
\begin{align}
\label{phasemodel-inertia}
m \ddot \theta_j + \dot \theta_j =  \omega_j + \frac{K}{N} \sum _{k=1} ^N \sin(\theta_k - \theta _j),
\end{align}
where $m$ corresponds to the strength of the inertia.
The inertia introduces an adaptive effect in the system \cite{Tanaka}.
Numerical simulations have confirmed that Eq.~$(\ref{criticaleq})$ holds even for large values of $m$, i.e., for strong adaptive effects \cite{Tanaka}.
Third, we explore the effect of amplitudes by using the globally coupled Stuart--Landau equations \cite{Matthews}, described as follows:
\begin{align}
\label{SL}
\dot z_j = (1 - |z_j|^2 + i \omega_j) z_j + \frac{K}{N} \sum _{k=1} ^{N} (z_k - z_j),
\end{align}
where $z _j\in \mathcal{C}$ represents the complex state variable of the $j$th oscillator.
The variance of the natural frequencies $\omega _j$ is assumed to be not so large;
otherwise, collective chaos or oscillation death occurs and violates Eq.~$(\ref{criticaleq})$ \cite{Matthews}.

\begin{figure}[h]
\begin{center}
 \includegraphics[width=5cm]{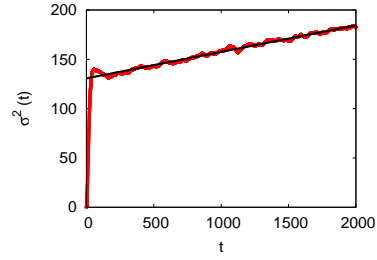}
 \caption{The time evolution of the variance $\sigma ^2 (t)$ of the integrated order parameter
in the coherent regime of system (\ref{SL}) with amplitude variables, where $N=8000$ and $K=0.31$.
The value of $\sigma ^2(t)$ increases linearly with slope $2Dt$ after a transient period.
The natural frequencies $\omega_j$ are randomly generated from a Gaussian distribution $G(\omega)$ with mean 0 and standard deviation 0.2.
The red line shows an average over 300 different initial conditions.}
 \label{fig:estimation}
 \end{center}
\end{figure}

\begin{figure}
\centering
\begin{center}
\begin{tabular}{lll}
\includegraphics[width=5.2cm,clip]{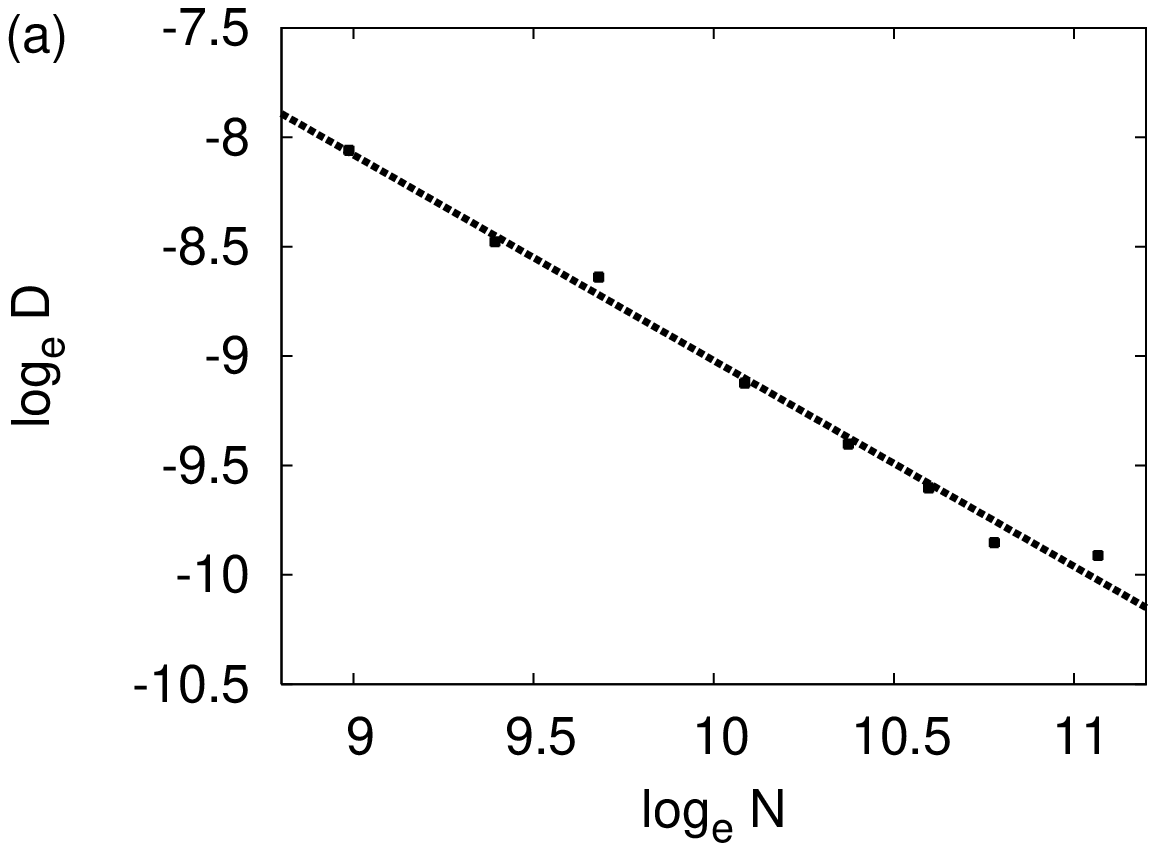}
\includegraphics[width=5.2cm,clip]{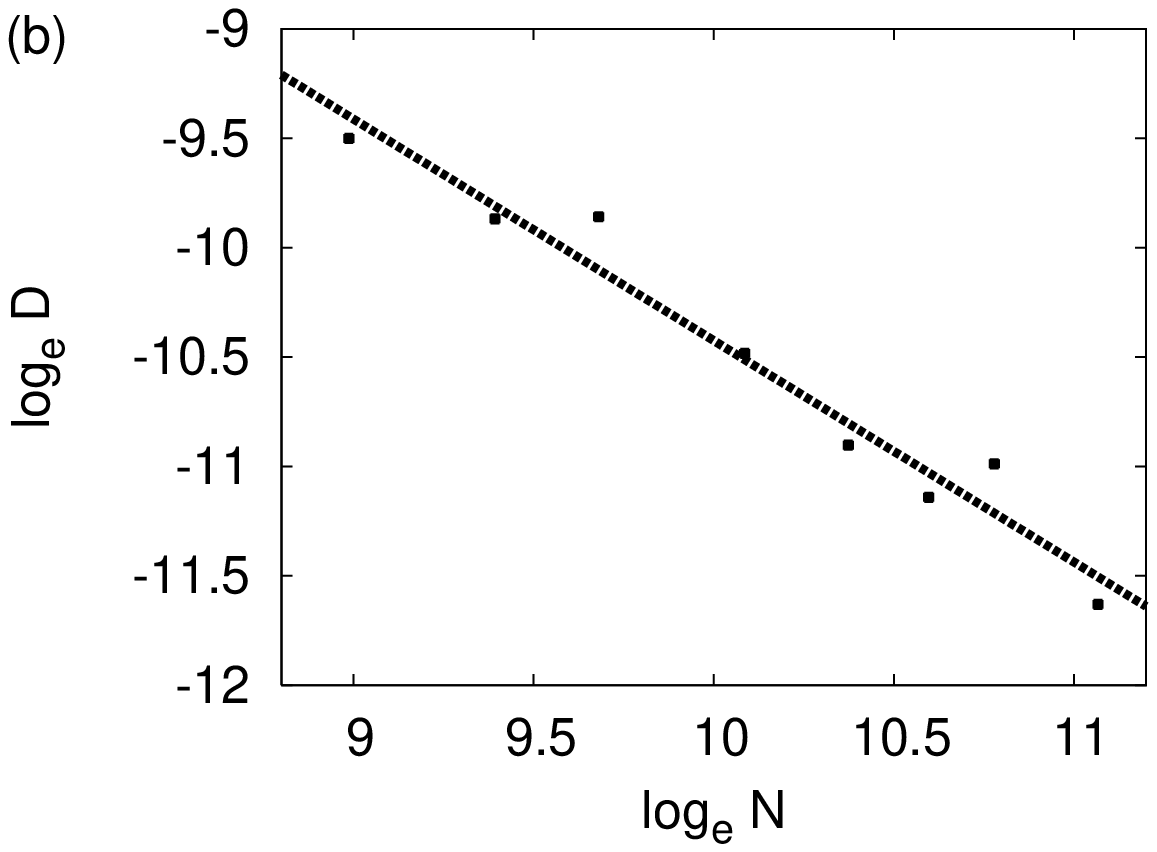}
\includegraphics[width=5.2cm,clip]{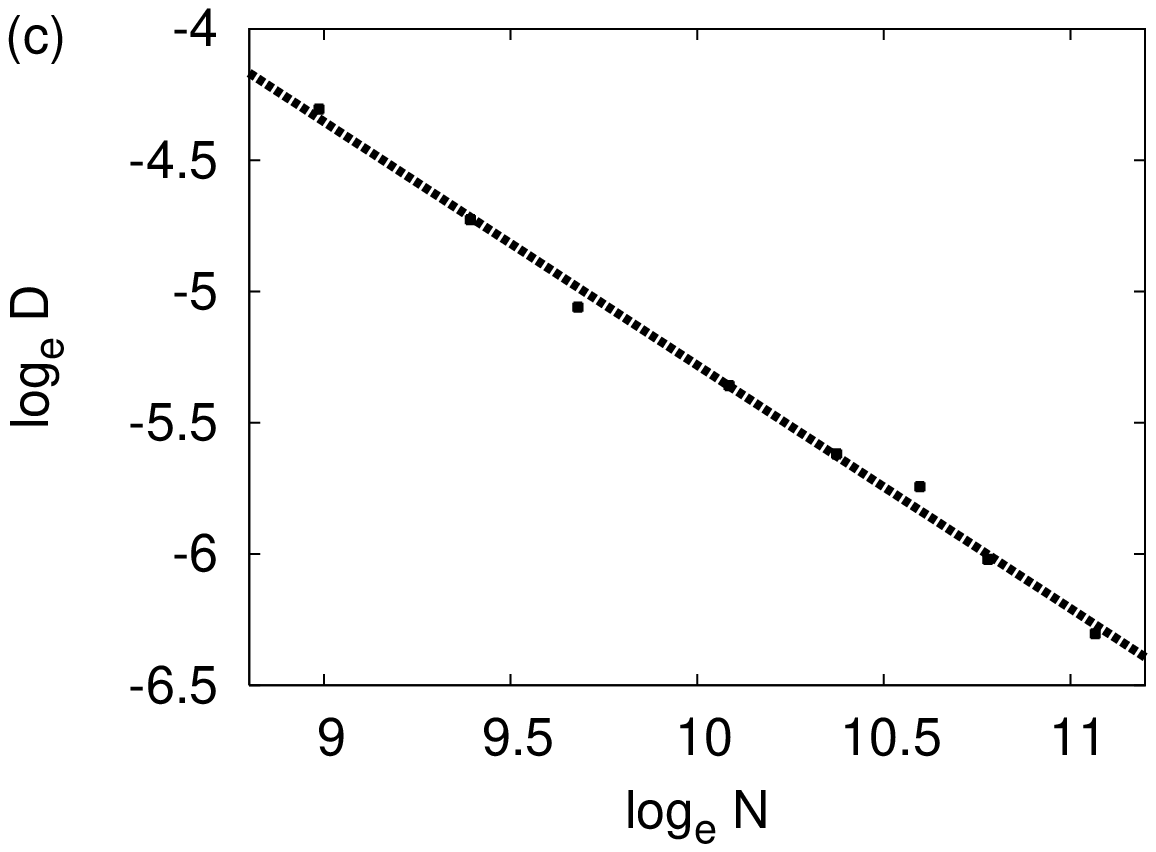}
\end{tabular}
\begin{tabular}{lll}
\includegraphics[width=5.2cm,clip]{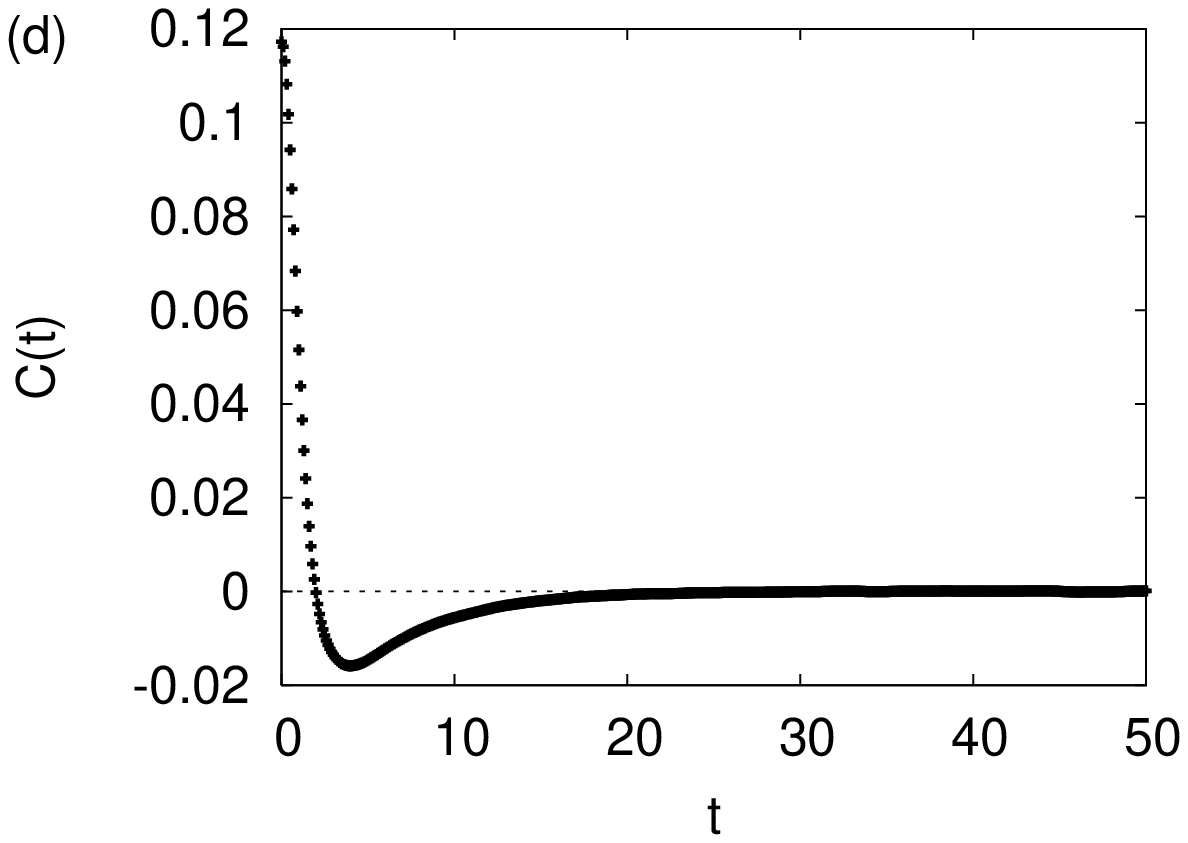}
\includegraphics[width=5.2cm,clip]{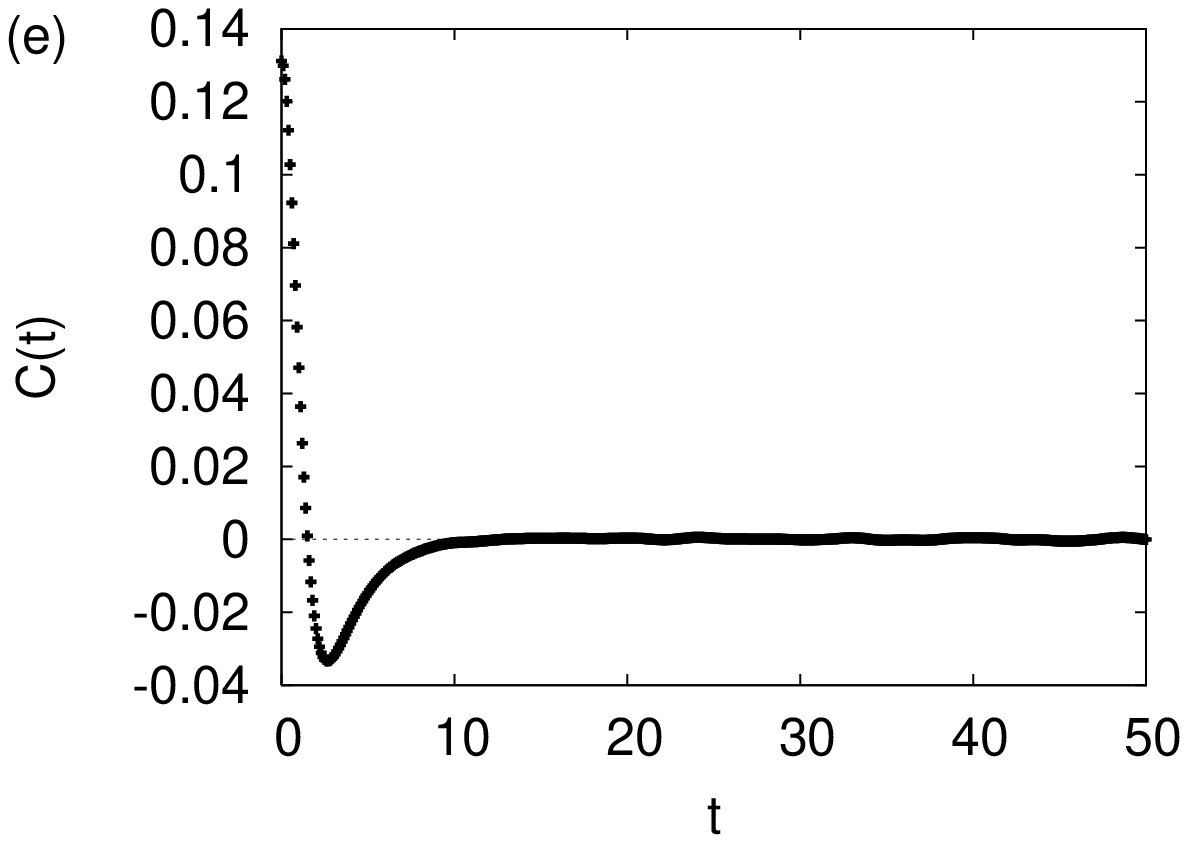}
\includegraphics[width=5.2cm,clip]{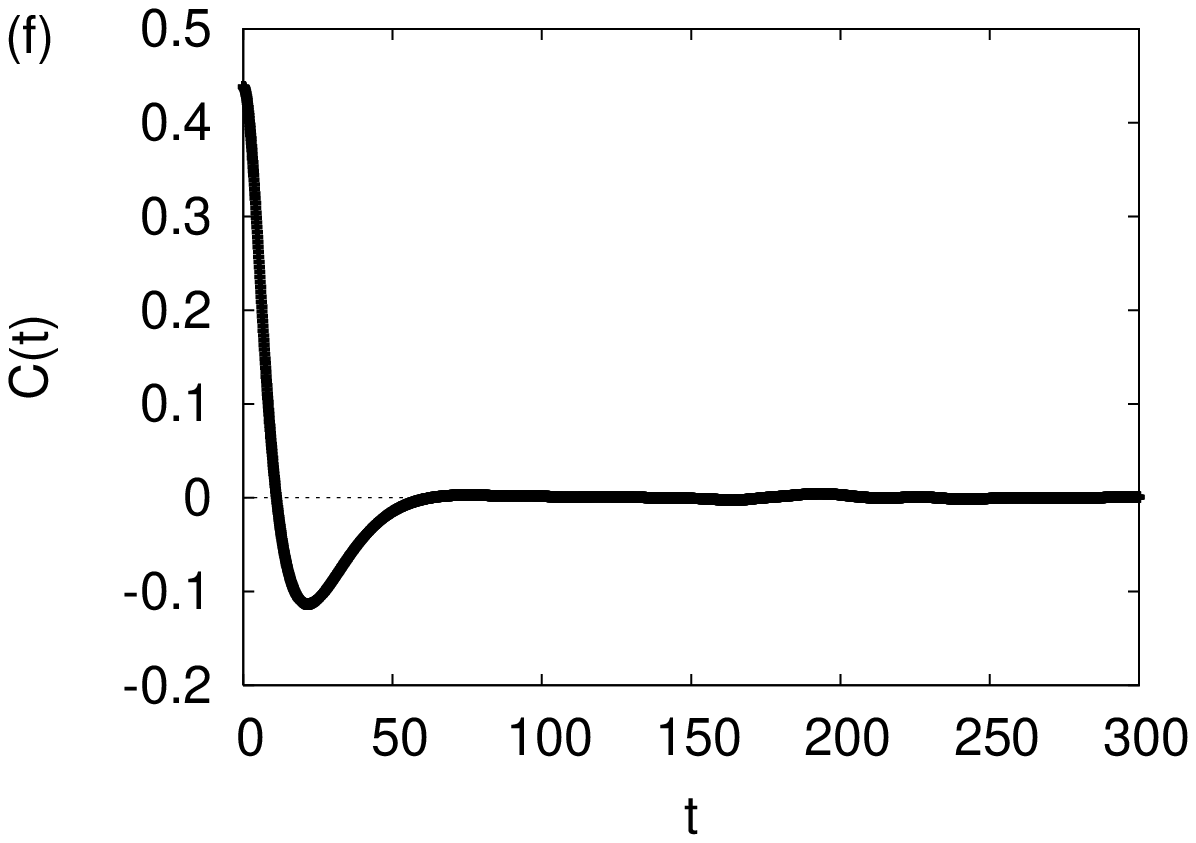}
\end{tabular}
\end{center}
\caption{Behavior of statistical quantities in the coherent regime.
The upper panels show the scaling properties of $D$ with respect to the system size $N$ ranging from 8000 to 64000.
The lower panels show the correlation function $C(t)$ for $N=64000$.
Each plot is an average over 100 different initial conditions.
The diffusion coefficient $D$ is scaled as $D \sim O(1/N^a)$ with $a \sim 0.940$ for (a), $a \sim 1.012$ for (b), and $a \sim 0.927$ for (c).
(a) and (d) System (\ref{phasemodel-timedelay}) with time delayed coupling, where time delay $\tau =1$ and $K = 1.78$.
We used the order parameter given by Eq.~(\ref{R}).
(b) and (e) System (\ref{phasemodel-inertia}) with inertia term, where $m=0.1$ and  $K = 1.93$.
We used the order parameter given by Eq.~(\ref{R}).
(c) and (f) System (\ref{SL}) with amplitude variables, where $K = 0.325$.
We used the order parameter defined as $R(t) \equiv \left| \sum _{j=1} ^N z _j \right|/N$.
The natural frequencies $\omega_j$ are randomly generated from a Gaussian distribution $G(\omega)$ with mean zero and a fixed standard deviation.
The standard deviation of the natural frequencies $\omega _j$ is set at 1 for systems (\ref{phasemodel-timedelay}) and (\ref{phasemodel-inertia}), and at 0.2 for system (\ref{SL}).
In the case where the natural frequencies are given deterministically i.e. $j/(N+1)=\int_{-\infty}^{\omega_j} G(\tilde{\omega})d\tilde{\omega}$,
the result does not change.}
\label{fig:DC}
\end{figure}

We show that all of these models satisfy the scaling law $(\ref{DN})$.
The value of $D$ is estimated by fitting $\sigma ^2(t)$ with a line of slope $2Dt$ except for the transient period,
as shown in Fig.~$\ref{fig:estimation}$.
For all the systems considered in the present paper,
we clearly see that $D$ is scaled as $D\sim O(1/N^a)$, as shown in Figs.~$\ref{fig:DC}$(a)-(c).
In addition, the correlation function has a characteristic time period in which $C(t) < 0$ as shown in Figs.~$\ref{fig:DC}$(d)-(f).
Therefore, it is suggested that
Eq.~$(\ref{criticaleq})$  is a sufficient condition for $D\sim O(1/N^a)$ in globally coupled oscillator systems with large $N$.

\section{Discussion}

We have investigated the condition under which
the non-standard scaling law $(\ref{DN})$ for the statistical quantity $D$ holds.
We have considered the relation between the non-standard scaling law and the condition in Eq.~$(\ref{criticaleq})$,
and we have performed numerical simulations of several models satisfying the condition.
The numerical results suggest that the condition is not only a necessary but also a sufficient condition 
for the non-standard scaling law.
Our non-standard scaling law $(\ref{DN})$ does not sensitively depend on details of coupling schemes \cite{Nishikawa}
as well as on other factors such as time delay, inertia, and amplitude variables as demonstrated in this study.
It is noteworthy that
such a non-standard scaling law that can be observed in a large class of globally coupled oscillator models
has not been reported thus far.
In fact, even for conventional statistical quantities such as the variance and the correlation time of the order parameter,
scaling laws with respect to coupling strength $K$ and those with respect to system size $N$
have not been fully explored except for those of the phase oscillator model (\ref{phasemodel})
\cite{PikovskyBook,Kuramotomodel,Daido,Pikovsky4,Hong,Hong3,Hildebrand,Buice,Nishikawa,Nishikawa2}.
Thus, our result is useful for understanding fluctuations in general systems of globally coupled oscillators.

\section*{Acknowledgments}
IN is grateful to the former supervisor, the late Prof. Hirokazu Fujisaka,
for his helpful discussions.
In fact, inspired by his paper (Fujisaka et al., Physica D (2005)),
IN conceived this work.
We dedicate this study to Prof. Fujisaka.
This research is supported by
Grant-in-Aid for Scientific Research (A) (20246026) from MEXT of Japan,
and by
the Aihara Innovative Mathematical
Modelling Project, the Japan Society for the Promotion of Science
(JSPS) through the ``Funding Program for World-Leading Innovative R\&D
on Science and Technology (FIRST Program)," initiated by the Council
for Science and Technology Policy (CSTP).


\clearpage

\clearpage

\clearpage

\end{document}